\documentstyle[12pt]{article}

\newcommand{\be}{\begin{equation}}
\newcommand{\ee}{\end{equation}}
\newcommand{\bea}{\begin{eqnarray}}
\newcommand{\eea}{\end{eqnarray}}

\begin{document}
\begin{titlepage}

%\flushright{IP-BBSR-2005-02 }
\begin{flushright}
{\today}
\end{flushright}
\vspace{1in}

\begin{center}
\Large
{\bf SCATTERING OF STRINGY STATES IN COMPACTIFIED CLOSED BOSONIC STRING}
\end{center}

\vspace{.5in}

\normalsize

\begin{center}
{ Jnanadeva Maharana \footnote{Raja Ramanna Fellow} \\
E-mail maharana$@$iopb.res.in} 
\end{center}

\normalsize
%\vspace{.1in}

\begin{center}
 {\em Institute of Physics \\
Bhubaneswar - 751005, India  \\
and \\
KEK Theory Center, Institute of Particle and Nuclear Studies,\\
KEK, Tsukuba, Ibaraki 305-0801, Japan  }

\end{center}

\vspace{.2in}

\baselineskip=24pt

\begin{abstract}
We present scattering of stringy states of closed bosonic string compactified
on torus $T^d$. We focus our attention on scattering of moduli and gauge
bosons. These states appear when massless excitations  such as graviton and
antisymmetric tensor field of the uncompactified theory are dimensionally
reduced to lower dimension. The toroidally compactified theory is endowed
with the $T$-duality symmetry, $O(d,d)$. Therefore, it is expected that  the
amplitude for scattering of such states will be $T$-duality invariant.
The formalism of Kawai-Llewelen-Tye is adopted and appropriately tailored
 to construct the vertex operators of moduli and gauge bosons. It is
shown, in our approach, that N-point amplitude is T-duality invariant. We
present illustrative examples for the four point amplitude
to explicitly demonstrate the economy of our
formalism when three spatial dimensions are compactified on $T^3$.
It is also shown that if we construct an amplitude with a set of
'initial' backgrounds, the $T$-duality operation transforms it to an 
amplitude associated with another
set backgrounds. 
We propose a modified version of KLT approach to construct vertex
operators for nonabelian massless gauge bosons which appear in certain
compactification schemes.

\end{abstract}

\vspace{.7in}

\end{titlepage}

\section{Introduction }

The perturbatively consistent bosonic and superstring theories live in
critical dimensions $26$ and $10$ respectively. The five superstring theories:
type IIA, type IIB, type I with $SO(32)$ gauge group, heterotic string
with gauge group $SO(32)$ and heterotic string with gauge group
$E_8\otimes E_8$ offer the prospect of unifying the fundamental forces 
\cite{book1,book2,book3}. The heterotic string \cite{het1,het2} is the most
promising candidate in the endeavors of unification programme and has
accomplished several notable successes. The ingenious construction of
heterotic string exploits an important attribute of closed string theory that
the left moving and right moving sectors are independent so far as the
evolution of the free string is concerned. Therefore, the left moving sector
is a closed bosonic string with critical dimension $26$ whereas the right
moving sector is a $10$-dimensional superstring. Of the $26$ bosonic 
coordinates $16$ are compactified on a torus, $T^{16}$, and the quantized
momenta along the compact directions are required to satisfy certain 
constraints. Consequently, the theory defined in $10$-dimensional
spacetime  manifests
$N=1$ supergravity and supersymmetric Yang-Mills multiplets in its massless
sector. The emerging gauge group, in the compactification scheme, is 
either $SO(32)$ or $E_8\otimes E_8$ depending on the boundary conditions
chosen for the compact coordinates. The underlying gauge groups are elegantly
unraveled once the compact bosonic coordinates are feminized and suitable
boundary conditions are assigned to the resulting Weyl Majorana fermions.
In order to establish connections with unified theories in four spacetime
dimensions, the heterotic string theory (and for that matter other superstring
theories) are to be reduced to four dimensional theories. This is achieved
by compactifying six spacial dimensions. One of the universal features of
the compactifications in string theory is the discovery of rich strove of
symmetries. One encounters new gauge symmetries as well as global symmetries.
The toroidal compactification has attracted considerable attention
over decades. One of the most interesting results in this programme is
the proposal of Narain \cite{narain} where he demonstrated the existence
of rich symmetry structure in toroidal compactification of heterotic string
and its salient features were further explored by Naranain, Sarmadi and Witten
\cite{nsw}. One of the central features in toroidal compactification is the
existence of noncompact target space duality symmetry (the T-duality).
One of the simplest example is to consider toroidal compactification of a
 closed bosonic string on a d-dimensional torus, $T^d$. 
Let us consider the
effective action associated with the massless states, graviton, antisymmetric
tensor and dilaton of the uncompactified theory. 
The effective action is dimensionally reduced following the Scherk-Schwarz
procedure where the background fields are assumed to be independent of
the compact coordinates. In the context of the string effective action, the
reduced action is expressed in a manifestly O(d,d) invariant form and the
generalization to heterotic string effective action in also incorporated
appropriately \cite{jmjhs}.
\\
The existence of the T-duality symmetry and the invariance of the reduced
effective action under T-duality has important implications (for reviews see
\cite{rev,rev1,rev2,rev3,rev4,reva,revb,revjm}). For example, if a
set of background correspond to a string vacuum, it is possible to generate
a new set of backgrounds, satisfying the equations motion, by judiciously
implementing duality transformations. In general the new set could
correspond to an inequivalent string vacuum. Indeed, this prescription has
been very useful to generate new backgrounds from a given set in string
cosmology \cite{meisv}, in stringy black hole physics \cite{youm}
 and so on.\\
The scattering matrix - S-matrix - plays a very important role in string 
theory.  S-matrix can be computed perturbatively within the first quantized
frame work of string theory. We associate a vertex operator with each state
of the string theory. The vertex operator is constrained by imposing the 
requirements of conformal invariance. Therefore, it is of interests to 
investigate consequences of T-duality in scattering of stringy states in a
compactified string theory. We ignore the winding modes throughout this
investigation. These are special characteristics of compactified strings
since the string can wind around compact coordinates due to its one
dimensional nature. Interactions of strings, taking into effects of winding
modes have interesting consequences; however, we consider the scenario 
where the states propagate in the noncompact spacetime dimensions.
\\
I have proposed \cite{jm4} a prescription to construct duality invariant
scattering amplitudes for the massless states (moduli and gauge bosons) of
compactified bosonic string. These massless states appear from the
dimensional reduction of the massless sectors such as graviton and
antisymmetric tensor field. The vertex operators are necessary to describe
the scattering of stringy states. 
 The
duality transformation, $O(d,d)$ generates a new set of backgrounds and
their corresponding vertex operators. We evaluate the amplitude for the
new backgrounds. I demonstrated how the two amplitudes, computed from
the two sets of vertex operators, (which are related by duality transformation)
are connected to each other. This result is analogous to the application of
$T$-duality symmetry in the context where starting from a given initial
background configuration one generates a new set of them. However, in the
process
of evaluating scattering amplitudes it has to be kept in mind that vertex
operators are constructed in the weak field approximation. 
In contrast, when the $T$-duality transformation is implemented to generate
new backgrounds, we use the M-matrix. The initial set of backgrounds is
the one which satisfy equations of motion (i.e. $\beta$-function equation).
On the other hand, when we construct the vertex operators associated with
these backgrounds, we invoke the weak field approximation. The background
is expanded around trivial one, for example the vertex operator for the
graviton/moduli is constructed in linearized approximation to metric and
corresponding equations result from requirement that it be $(1,1)$ with
respect to free stress energy momentum tensor.
Therefore,
in my earlier
formulation, it is not a straight forward implementation of $T$-duality as is
the scenario adopted for duality transformation
exploiting  the solution generating technique.\\
Recently Hohm, Sen and Zwiebach (HSZ) have revisited the duality symmetry of
heterotic string effective action \cite{hsz}.
They analyze which duality symmetries are
realized in the effective field theory when the nonabelian gauge field sector
is kept without truncating it to the Cartan (Abelian) sector. In this context
these authors make some important observations about the attributes of
the tree level S-matrix of the moduli in order to draw conclusions about
some crucial features of S-matrix. Furthermore, based on their arguments,
they discuss properties of the effective action. They focus their attention
on the case when the
background fields are independent of the compact coordinates.
This is also one essential ingredients
of the our assumption in the present investigation and the same assumption
was invoked in \cite{jm4}. Furthermore, HSZ justify the reasons for focusing
their consideration
to  the massless sector in order to achieve their goals.
Therefore, the excitations of heterotic string, relevant for their purpose,
 are
the moduli and gauge bosons. Furthermore, they 
 provide arguments to substantiate the
$T$-duality invariance of the S-matrix.\\
We mention, {\it en passant}, that our principal goal is to study $T$-duality
invariance of S-matrix. As alluded to earlier, the intent is to construct
vertex operators which will lead us to compute amplitudes of  these massless
states for compactified bosonic string, in contrast to those of the
compactified heterotic string. We cast the amplitude, appealing to the newly
proposed vertex operators, in a $T$-duality invariant form. Our
explicit calculations are more efficient and economical than  earlier
techniques of mine \cite{jm4} and it lends supports to
the arguments of HSZ that S-matrix is
$T$-duality invariant.\\
In this article
we evaluate the tree level amplitudes to bring out essential features
of duality symmetry in the present context and study how T-duality acts on 
S-matrix elements. The string perturbation theoretic
corrections are not expected to affect duality transformation properties
of S-matrix.
\\
The rest of the article is organized as follows. In Section II, we recapitulate
essential features of compactified closed bosonic string from the worldsheet
perspective. We introduce the vertex operators associated with different 
levels of compactified string. Next we recall how the vertex operators could
be cast in $O(d,d)$ invariant form. The vertex operators are defined in the
weak field approximation in the sense that all correlation functions are
evaluated using the free field OPE in the computation of S-matrix
elements. In the next section, Section III, we resort to the prescription of
Kawai, Llewellen and Tye \cite{klt} for computation of S-matrix elements. We
argue that duality transformation properties of the vertex operators become
rather transparent in KLT formulation. Moreover, this technique  is also 
very useful when
we want to study duality properties of excited massive states. We construct
N-point amplitude of moduli. These massless scalar appear when the metric
and antisymmetric tensor field, massless states of the closed string, are
compactified to lower dimensions. The massless Abelian gauge bosons also
appear as a consequence of toroidal compactification. We also provide 
expressions for N-point amplitude of gauge bosons. Our principal goal is
to construct $T$-duality invariant amplitudes. To this end, we resort to
the proposal of Sen \cite{sen1}. 
In this section explicit duality invariant amplitude
is constructed. Moreover, we present illustrative examples of four point
functions where a given amplitude is shown to get related to another
amplitude through duality transformation. We propose a prescription to
construct N-point amplitude for nonabelian gauge bosons and nonabelian 
massive string
excitations in the spirit of KLT formalism. We discuss our results and
present conclusions in Section IV. 

\section{Duality Symmetry of Closed String and the S-matrix}
Let us consider the closed string worldsheet effective action in the
presence of constant metric and antisymmetric tensor field
\be
\label{bosonic}
S=-{1\over 2}\int d\sigma d\tau\bigg(\partial_a X^{\hat\mu}\partial^aX^{\hat\nu}
{\hat G}^{(0)}_{{\hat\mu}{\hat\nu}}+
\epsilon^{ab}\partial_a{\hat X}^{\hat\mu}\partial_b{\hat X}^{\hat\nu}
{\hat B}^{(0)}_{{\hat\mu}{\hat\nu}}
\bigg)
\ee
Here $\tau ,\sigma$ are the worldsheet coordinates, and 
${\hat\mu},{\hat\nu}=0,1,2.....{\hat D}-1$ and $\hat D$ is the number of
spacetime dimensions. We have already adopted orthonormal gauge for the
worldsheet metric. We have omitted $\alpha'$ from the definition of action
(\ref{bosonic}). The canonical Hamiltonian density is \cite{m1,m2}
\be
\label{hamiltonian}
H={\bf Z}^T{\cal M}_0{\bf Z} 
\ee
where
\be
\label{def}
{\bf Z}=\pmatrix{{{\hat P}_{\hat \mu}} \cr {{\hat X}'^{\hat\mu}}},~~
{\cal M}_0=\pmatrix{{\hat G}^{-1} & -{\hat G}^{-1}{\hat B} \cr
{\hat B}{\hat G}^{-1} & {\hat G}-{\hat B}{\hat G}^{-1}{\hat B}}
\ee
where ${\hat P}_{\hat \mu}$ is canonical conjugate momenta of ${\hat X}^{\hat\mu}$ and
prime stands for $\sigma$ derivative.
Note that ${\cal M}_0$ is a symmetric $2{\hat D} \times 2{\hat D}$ matrix. The
Hamiltonian is invariant under a global $O({\hat D}, {\hat D})$ symmetry
transformation
\be
\label{oddtrans}
{\bf Z}\rightarrow \Omega_0{\cal M}_0\Omega_0^T, ~~~\Omega_0^T{\hat \eta}\Omega_0
={\hat \eta}
\ee
where $\Omega_0\in O({\hat D}, {\hat D}) $ and ${\hat \eta}$ is 
the $O({\hat D}, {\hat D})$ metric, with zero 
diagonal elements and off diagonal elements being ${\hat D}\times {\hat D}$
unit matrices. If we consider evolution of string in the background of its 
massless excitations such as ${\hat G}$ and $\hat B$, they are allowed to have
${\hat X}^{\hat\mu}(X(z,{\bar z}))$ 
dependence and the resulting 
action is that of a sigma model. In toroidal compactification prescription
the ${\hat X}^{\hat\mu}$ is decomposed into two sets of coordinates
${\hat X}^{\hat\mu}=\{X^{\mu}, Y^{\alpha}\}, ~\mu=0,1...D-1,~
{\rm and}~ \alpha=D,D+1,..
{\hat D}-1$ so that ${\hat D}=D+d$; $\{Y^{\alpha} \}$ are the compactified
coordinates. Furthermore, the backgrounds depend only
on $X^{\mu}$. We shall summarize a few aspects which will be useful later. The
general prescription of Scherk and Schwarz \cite{ss} is the essential
ingredient of such compactifications. It is useful to adopt the vielbein
formalism 
\bea
\label{viel}
 {\hat { e}^{\hat A}_{\hat M}}({\hat X}) = 
\pmatrix{ e^{r}_{\mu}(X) &  A^{(1)\beta}_{\mu}(X)E^{a}_{\beta}(X) \cr
0 & E^{a}_{\alpha}(X) } 
\eea
The spacetime metric is $g_{\mu\nu}= e^{r}_{\mu}e_{\nu r}$ where the local
indices $r,s$ are raised and lowered by flat Minkowski metric. Note the 
appearance of Abelian gauge fields $A^{(1)\alpha}_{\mu}, \alpha=1,2...d$ 
associated with the $d$ isometries and $E^{a}_{\alpha}$ are vielbein of
the internal metric i.e. $G_{\alpha\beta}
= E^a_{\alpha}E^b_{\beta}\delta_{ab}$ and it transforms as a scalar under
$D$-dimensional Lorentz transformations. Similarly, the antisymmetric 
tensor will
be decomposed into components: ${\hat B}_{\mu\nu},~{\hat B}_{\mu\alpha},~
{\hat B}_{\alpha\beta}$. There are gauge fields ${\hat B}_{\mu\alpha}$ and
scalars ${\hat B}_{\alpha\beta}$ in lower dimensions. We refer the interested
reader to our paper \cite{jmjhs} for details and the prescriptions for 
dimensional reduction in the context of string theory. It is worth noting that
under the T-duality transformation the spacetime tensors remain invariant.
If we were to dimensionally reduce the $\hat D$ dimensional effective action
then all backgrounds depend only on $D$-dimensional spacetime coordinates
$x^{\mu}$. Moreover, the M-matrix defined bellow, is expressed in terms of
the moduli $G$ and $B$
\bea
\label{mmatrix}
M=\pmatrix{ G^{-1} & -G^{-1}B \cr BG^{-1} & G-BG^{-1}B }
\eea
where $G=G_{\alpha\beta}, ~ B=B_{\alpha\beta}$ and now $B_{\alpha\beta}$
 stands for ${\hat B}_{\alpha\beta}$. The $O(d,d)$ metric $\eta$ with
off diagonal $d\times d$ unit matrix, $\bf 1$ remain invariant. Under 
the $O(d,d)$ transformations
$M\rightarrow \Omega M\Omega ^T,~ \Omega \in O(d,d)$. $M$ is symmetric and
$M\in O(D,d)$. The gauge fields,
$A^{(1)\alpha}_{\mu}$ and $A^{(2)}_{\mu\alpha}={\hat B}_{\mu\alpha}+
B_{\alpha\beta}A^{(1)\beta}_{\mu}$
transform as a vector under $O(d,d)$ transformations i.e.
${\cal A}^i_{\mu}=\Omega ^i_j{\cal A}^{j}_{\mu},~ i,j=1,2...2d$. 
We define ${\cal A}^{i}_{\mu}=A^{(1)\alpha}_{\mu},~i=1,2...d$ and 
${\cal A}^{i}_{\mu}=A^{(2)}_{\mu\alpha}, ~ i=d+1,...2d$; note $\alpha$ takes
$d$-values. We shall need this transformation property of the gauge field in
sequel.\\
The vertex operators, $V(k;X(z,{\bar z}))$, in closed string theory are 
required two satisfy two constraints: 
\be
\label{vertex1}
(L_0-1)V=0,~~~({\tilde L}_0-1)V=0, ~~~{\rm and}~~~(L_0-{\tilde L}_0)V=0
\ee
The last constraint is the level matching condition. Before proceeding,
let us look at the well known graviton vertex operator
\be
\label{graviton}
V_g(\epsilon, k, X)=\epsilon_{\mu\bar\nu}
:e^{ik.X(z,{\bar z})}\partial X^{\mu}(z)
{\bar\partial}X^{\bar\nu}({\bar z}):
\ee
The above constrains (\ref{vertex1}) lead to two conditions: (i) the graviton
is massless, $k^2=0$ and (ii) it is transverse $\epsilon_{\mu\bar\nu}k^{\mu}=
\epsilon_{\mu\bar\nu}k^{\bar\nu}=0$. The graviton coupling 
$g_{\mu\bar\nu}\partial X^{\mu}
{\bar\partial}X^{\bar\nu}$ is given by (\ref{graviton}) 
in the plane wave approximation 
where $g_{\mu\nu}$ is expanded around flat background metric. 
We can adopt the same
prescription for the vertex operators associated with the moduli 
$G_{\alpha\bar\beta}~{\rm and}~ B_{\alpha\bar\beta}$. The corresponding vertex 
operators are
\be
\label{modulivertex}
V_G=\epsilon^G_{\alpha\bar\beta}:e^{ik.X(z,{\bar z})}\partial Y^{\alpha}(z)
{\bar\partial} Y^{\bar\beta}({\bar z})
\ee
Here $\epsilon^G_{\alpha\bar\beta}$, symmetric under 
$\alpha \leftrightarrow \bar\beta$, 
is analog of polarization tensor carrying indices along internal direction.
Moreover, the scalar propagates only in spacetime manifold and therefore,
the vertex operator has no dependence on $Y^{\alpha}$ and its (quantized)
canonical momenta $P_{\alpha}$. Thus it implies that
 we are not taking into account the presence of
winding modes. A similar vertex operator $V_B$ can be constructed for the other
moduli, B, in weak field approximation. 
Our aim is to study T-duality invariance
properties of the S-matrix for states of compactified closed bosonic string.
 Although transformation properties of
M-matrix  under T-duality are easy to implement (see 
the  remarks after (\ref{mmatrix})), the transformation properties 
 of the moduli $G$ and $B$ are rather complicated \cite{jmjhs}. In fact
$G+B$ transform as quotients under $O(d,d)$.
I have proposed \cite{jm4}
a prescription to evaluate amplitudes in terms of M-matrix elements in the 
weak field approximation. This proposal can be implemented in the massless
sector; in other words we can compute amplitudes involving the moduli
and generate new amplitudes through T-duality transformations. The procedure
consists of following steps.  Let us start with the vertex operator
(\ref{modulivertex}) and examine how T-duality operation works in my scheme.
(i) Just as we expand backgrounds $G_{\alpha\bar\beta}$ and 
$B_{\alpha\bar\beta}$ 
around trivial background, I propose such an expansion for the $M$-matrix:
$M={\bf 1}+{\tilde M}$ in weak field expansion. Since $M$ is expressed in terms
of $G$ and $B$, ${\tilde M}$ will be expressed in terms of the linearized
expansion of those backgrounds. Therefore, ${\tilde M}$ is constructed
 in terms of  $\epsilon^G_{\alpha\bar\beta}~ {\rm and} ~ 
\epsilon^B_{\alpha\bar\beta}$.
However, unlike $M$ which is an element of $O(d,d)$, 
${\tilde M}$ is not an $O(d,d)$ matrix. Consequently,
we cannot implement T-duality transformation directly on $\tilde M$.
\\
(ii) Thus my starting point was to  identify the vertex operators associated
with the moduli. I utilized $\epsilon^G$ and $\epsilon ^B$ to construct
$\tilde M$ and then the matrix $M={\bf 1}+{\tilde M}$.
\\
(iii) Now we can implement the $O(d,d)$ transformation on the constructed
$M$-matrix, i.e
$M\rightarrow ~ M'=\Omega M\Omega^T$ where $\Omega\in O(d,d)$.\\
(iv) The next step is to expand $M'$ as $M'={\bf 1}+{\tilde M}'$. Finally,
we can extract transformed
$\epsilon'^G_{\alpha\bar\beta}$ and $\epsilon'^B_{\alpha\bar\beta}$ from 
${\tilde M}'$.
\\
However,
for compactified string, the T-duality transformation is expected to operate
for massive levels arising from compactifications which have their 
partners as spacetime tensors at the same excitation level of the string. 
Indeed, I introduced a technique to construct $O(d,d)$ invariant vertex
operators arising from toroidal compactification of the closed string 
\cite{jm1}.\\
Let us look at a generic vertex operator \cite{jm2} which appears after a 
$\hat D$-dimensional vertex operator has been compactified to $D$ dimensions.

\bea
\label{generic}
{\partial_+}^pY^{\alpha_i}{\partial_+^q}Y^{\alpha_j}{\partial_+^r}
Y^{\alpha_k}..
{\partial_-}^{p'}Y^{\alpha'_i}\partial_-^{q'}
Y^{\alpha'_j}{\partial_-}^{r'}Y^{\alpha'_k}.,~~p+q+r=p'+q'+r'=n+1
\eea
where $\partial_{\pm}=\partial_{\tau}\pm \partial_{\sigma}$. Thus 
$(\partial_{\tau}\pm \partial_{\sigma})Y^{\alpha}=(P^{\alpha}\pm Y'^{\alpha})$.
The indices $\alpha,\beta..$ are raised and lowered by $\delta^{\alpha\beta}$
and $\delta_{\alpha\beta}$. The constraint $ p+q+r=n+1,~p'+q'+r'=n+1$ is 
required by the 
level matching condition. It is possible to express (\ref{generic}) as an
$O(d,d)$ tensor by introducing projection operators \cite{jm2} which convert 
$\partial_+Y$ and various products of $(\partial_+)^nY$ appearing in
(\ref{generic}) to products of $O(d,d)$ vectors. The same argument
applies to $(\partial{_-})^{m'}Y$ as well. 
Once  (\ref{generic}) is converted to products of $O(d,d)$ vectors, it has
to be contracted with corresponding $O(d,d)$ polarization tensor. This 
prescription can be implemented for first couple of massive levels. However,
even the formal expression for amplitudes involving such vertex operators
is not easy to manipulate to study T-duality invariance properties of
the S-matrix elements. 
Therefore, it is necessary to adopt a different strategy.\\
There is a proposal due to Sen \cite{sen1}, based on string field
theory, according to which  the space of solutions of backgrounds enjoys an
$O(d)\otimes O(d)$ symmetry. He noted that the diagonal subgroup, $O(d)$, of
$O(d)\otimes O(d)$ generates rotations. Furthermore, he identifies the
set of matrices
\be
\label{senod}
\Omega_{RS}={1\over 2}\pmatrix{S+R & R-S \cr R-S & S+R \cr }
\ee
which implement $O(d)\otimes O(d)$ transformations. 
Here $R$ and $S$ are matrices
that belong to $O(d)\otimes O(d)$ and $\Omega_{RS}$ is subgroup of $O(d,d)$. In the 
linearized approximation to the backgrounds
\be\label{linearapr}
G_{\alpha\beta}=\delta_{\alpha\beta}+h_{\alpha\beta},~~{\rm and}~~
B_{\alpha\beta}=b_{\alpha\beta}
\ee
the transformed linearized backgrounds are
\be
\label{gbtrans}
(h'+b')=S(h+b)R^T
\ee
This argument has been advanced further by Sen \cite{sen2} and by 
Hassan and Sen 
\cite{hs1,hs2} and the matrix defines in (\ref{senod})  operates on
transformations of the  $M$-matrix since $\Omega_{RS}\in O(d,d)$. \\
In order to evaluate scattering amplitudes involving moduli $G$ and $B$, we
employ 
the vertex operator (\ref{modulivertex}). The tree level N-point amplitude
is evaluated by utilizing the conformal field theory prescription. We have
 noted that two sets of Abelian gauge fields, $A^{(1)\alpha}_{\mu}$
and $A^{(2)}_{\mu\alpha},~ \alpha=1,2... d$ also appear in general toroidal
compactification scheme. Note that 
 gauge fields do not appear in compactification
proposed by Hassan and Sen \cite{hs1} 
since the backgrounds ${\hat G}$ and $\hat B$ are
decomposed in the block diagonal form. These two sets of gauge fields combined
together transform 
as $O(d,d)$ vectors as already noted. Therefore, the transformation
rules for these gauge bosons, under Sen's $O(d)\otimes O(d)$ group,
can be specified easily from the 
structure of $\Omega_{RS}$ matrix. We shall exploit this information when
we return to discussion of the scattering of these gauge bosons in sequel.
The N-point amplitude for the scattering
of moduli is
\be
\label{npoint}
A^{(N)}_{G,B}=\int d^2 z_1d^2z_2..d^2z_N{\bf <}{\bf\Pi}_{i=1}^NV_i(\epsilon_i,
k_i,X_i,Y_i){\bf >}
\ee
where the vertex operator $V_i$ is
\be
\label{vertexgb}
V_i(\epsilon_i,k_i,X_i,Y_i)=\epsilon_{{\alpha_i}{\bar\beta_i}}:{\rm exp}
[ik_i.X(z,{\bar z})]\partial Y^{\alpha_i}(z){\bar\partial}
Y^{\bar\beta_i}({\bar z})
:
\ee
with $k_i^2=0$. Where $\epsilon_{{\alpha_i}{\bar\beta_i}}$ 
stands for polarization
tensor of $G_{{\alpha_i}{\bar\beta_i}}$ or $B_{{\alpha_i}{\bar\beta_i}}$
 depending
on the choice we make and then it will be symmetric or antisymmetric under
$\alpha_i \leftrightarrow \bar\beta_i$. 
Note that the plane wave part, ${\rm exp}[
ik_i.X_i]$, is inert under T-duality and we shall not bring in its presence
in our considerations of duality symmetry transformations. While evaluating
the above amplitude (\ref{npoint}), we have to insert the Koba-Nielsen factor
and the integration is to be carried out on $N-3$ variables. We shall pay
attention to these aspects when we compute amplitudes for specific cases.
However, we remind the reader about the two correlation functions which we
shall use from time to time.
\be
\label{correl1}
<\partial Y^{\alpha_i}(z_i)\partial Y^{\alpha_j}(z_j)>=-
{{\delta^{{\alpha_i}{\alpha_j}}}\over{(z_i-z_j)^2}},
\ee
and
\be
\label{correl2}
<{\bar\partial} Y^{\bar{\beta}_i}({\bar z}_i)
{\bar\partial} Y^{{\bar\beta}_j}({\bar z}_j)>=
-
{{\delta^{{{\bar\beta}_i}{{\bar\beta}_j}}}\over{({\bar z}_i-{\bar z}_j)^2}}
\ee
We have used $\alpha'=2$ for close string through out since all
the amplitudes involve the vertex operators associated with the closed
string states. Therefore, it does
not appear in our computations. If we have to introduce it then the two
correlation functions will be multiplied by a factor ${\alpha'}\over 2$ on
the right hand sides of equations (\ref{correl1}) and (\ref{correl2}). 
When the need will arise we shall remind where the $\alpha'$ be introduced.
We draw in discussions about vertex operators of open string in the 
next section very briefly in the context of KLT formalism. We have suppressed
the $\alpha'$ factors in this context also.
Thus the products of $\epsilon_{{\alpha_1}{\bar\beta_1}},
\epsilon_{{\alpha_2}{\bar\beta_2}},...\epsilon_{{\alpha_N}{\bar\beta_N}}$  
get contracted
with various combinations of 
$\delta^{{\alpha_i}{\alpha_j}}, \delta^{{\bar\beta_i}{\bar\beta_j}},...$ which will
come from pairwise contractions of holomorphic parts and antiholomorphic parts
the the vertex operators. The plane wave contractions are like
$<:{\rm exp}[ik_i.X_i(z_i,{\bar z}_i)]::{\rm exp}[ik_j.X_j(z_j,{\bar z}_j)]:>
=|z_i-z_j|^{2k_i.k_j}$. If we adopt Sen's prescription of $O(d)\otimes O(d)$
transformations then
\be
\label{trans1}
(\epsilon'^G_{{\alpha_i}{{\bar\beta}_i}}+ 
\epsilon'^B_{{\alpha_i}{{\bar\beta}_i}})=
[S(\epsilon^G+\epsilon^B)R^T]_{{\alpha_i}{{\bar\beta}_j}}  
\ee
Let us consider the amplitude for scattering of gauge bosons(we confine to
amplitude for $A^{(1)\alpha_i}_{\mu_i}$ for the moment).
\be
\label{gaugeA}
T^{(N)}_A=\int d^z_1... d^2z_N{\bf <}
{\bf \Pi}_{i=1}^NV^A_i(\epsilon_1,k_i,X_i,Y_i){\bf >}
\ee
where
\be
\label{gaugevertex}
V^A_i(\epsilon_i,k_i,X_i,Y_i)=
\epsilon_{{\mu_i}{\bar\alpha_i}}:
{\rm exp}[ik_i.X(z_i,{\bar z}_i)]\partial X^{\mu_i}
(z_i){\bar\partial}Y^{\bar\alpha_i}({\bar z}_i)
\ee
and  is required to satisfy constraints: (i) $k_i^2=0$ and 
(ii)$\epsilon_{{\mu_i}{\bar\alpha_i}}k^{\mu_i}=0$. Thus contracted terms
 in the amplitude will be multiplied by products of 
$\epsilon_{{\mu_i}{{\bar\alpha}_i}}$ which will be contracted by 
various combinations spacetime metric and internal metric.\\
We would like to draw attention to the fact that $(\epsilon^G+\epsilon^B)
\rightarrow S(\epsilon^G+\epsilon^B)R^T$ under $O(d)\otimes O(d)$ 
transformations and $\epsilon^{(1)}_{{\mu_i}{\bar\alpha_i}}\rightarrow
(S+R) \epsilon^{(1)}_{{\mu_i}{\bar\alpha_i}}$. In the expression for the 
N-point amplitude for the scattering of moduli (\ref{npoint} ) we encounter
a string of the products of $\epsilon^G$ and $\epsilon^B$. Therefore, it is
not yet so straight forward to demonstrate the T-duality invariance of the
amplitude although Sen's argument of implementing $O(d)\otimes O(d)$
duality is more efficient compared to my earlier proposal.
It turns out that the technique introduced by Kawai, Llewellen and Tye
\cite{klt} is very useful to investigate the T-duality transformation
properties of scattering amplitude in the frame work of Sen's 
$O(d)\otimes O(d)$ duality group.

\section{KLT Formalism and T-duality}
Let us briefly recall salient features of the technique introduced by KLT
\cite{klt} to calculate tree level closed string amplitudes. We shall
appropriately modify their prescription for our purpose. Their principal
goal was to demonstrate the relationship between closed string and open string
tree level scattering amplitudes. They showed that the N-point closed string
amplitude can be factorized into products of two N-point open string amplitudes
with certain pre-factors. This property holds for all excited levels of string
theories as long as one confines to the vertex operators associated with
states lying on the leading Regge trajectory. It was observed that closed 
string coordinates are decomposed as sum of left and right moving sectors.
The vertex operator corresponding to leading trajectory is product of
 equal number of holomorphic and antiholomorphic operators i.e
one set is ${\bf \Pi}_{i=1}^M\partial X^{\mu_i}(z_i)$ and the other one is 
 ${\bf \Pi}_{i=1}^N{\bar\partial} X^{\nu_i}({\bar z}_i)$; further more at each
level $M=N$ in order to satisfy the {\it level-matching} condition. 
The factorization property of three point and four point amplitudes was
explicitly demonstrated. In order to make
this article self contained, we summarize essential steps prescribed by KLT.
We shall not utilize the entire mechanism of KLT; however, as we shall
demonstrate, their approach brings out certain simplifications in actual
computations. It is also quite relevant in the present context to 
investigate $T$-duality invariance of the S-matrix.\\
Let us consider below the  open string vertex operators for a tachyon and a 
gauge boson respectively.
\be
\label{tg}
V^T(k,X)=:{\rm exp}[ik.X]:,~~{\rm and} ~~ V^A=\epsilon_{\mu}:{\rm exp}[ik.X]
\partial X^{\mu}: 
\ee
The tachyon and gauge boson are required to be on-shell; 
moreover,  the gauge boson polarization vector 
is transverse i.e. $k.\epsilon=0$. The scattering amplitudes involving
tachyons and gauge bosons are evaluated by adopting known techniques. 
Furthermore,
we need to introduce vertex operators with each excited level of the string.
KLT introduced an ingenious and elegant technique derive the scattering
amplitudes for excited level through modification of the tachyon vertex 
operator as follows. Consider the vertex operator
\be
\label{kltvert}
V^{open}_{KLT}(\epsilon,k,X)=:{\rm exp}[ik.X+i\epsilon_{\mu}\partial X^{\mu}]:
\ee
Now if we expand the exponential in powers of $\epsilon_{\mu}$, the linear
term in the polarization vector reproduces gauge boson vertex operator.
Moreover, if we desire to compute scattering amplitude for N gauge bosons,
$T_A^{(N)}$, then we compute the correlation function of products of $V_{KLT}$
and isolate the coefficient of the product ${\bf \Pi}_{i=i}^N\epsilon_{\mu_i}$.
Furthermore, the vertex operator for generic excited level of open string
is
\be
\label{excited}
V^{EX}\simeq\epsilon_{\mu_1\mu_2....\mu_m}:{\rm exp}[ik.X]
\partial X^{\mu_1}\partial X^{\mu_2}...\partial X^{\mu_m}:
\ee
The excited state momentum $k_{\mu}$ has to be on-mass-shell  and the
polarization tensor, $\epsilon_{\mu_1\mu_2....\mu_m}$, is required to satisfy
some transversality, tracelessness conditions as a consequence of conformal
invariance. According to KLT 
\be
\label{kltex}
V^{EX}_{KLT}\simeq :{\rm exp}[i(k.X+\epsilon_{\mu_1}\partial X^{\mu_1}+
\epsilon_{\mu_2}\partial X^{\mu_2}...+\epsilon_{\mu_m}\partial X^{\mu_m})]:
\ee
Now by expanding the exponential and  keeping the multilinear term
\be
\label{multi}
\epsilon_{\mu_1}\partial X^{\mu_1}\epsilon_{\mu_2}\partial X^{\mu_2}...
\epsilon_{\mu_m}\partial X^{\mu_m}
\ee
we recover the desired vertex operator (\ref{excited}).\\
The procedure outlined for open string is also generalized for the vertex
operator of closed string states. We recall that the graviton vertex operator
is
\be
\label{grav}
V^{graviton}=\epsilon_{\mu\bar\nu}:{\rm exp}[ik.X]\partial X^{\mu}
{\bar\partial}X^{\bar\nu}:
\ee
with $k^2=0$ and $\epsilon_{\mu\bar\nu}k^{\mu}=0=\epsilon_{\mu\bar\nu}k^{\nu}$.
 The
corresponding vertex operator for antisymmetric tensor assumes a similar form,
only exception being that $\epsilon_{\mu\bar\nu}$ is antisymmetric. 
The KLT prescription for generating vertex operator the massless sector of 
closed string is to introduce
\be
\label{kltclosed}
V^{closed}_{KLT}=:{\rm exp}[ik.X+i\epsilon_{\mu}\partial X^{\mu}+
i{\bar\epsilon}_{\bar\nu}{\bar\partial}X^{\bar\nu}]:
\ee
If we collect the coefficient of the bilinear term $\epsilon_{\mu}
{\bar\epsilon}_{\bar\nu}$ in the expansion of the exponential
(\ref{kltclosed}) we note
that it corresponds to graviton and antisymmetric tensor vertex if we
identify the symmetric product of $\epsilon_{\mu}{\bar\epsilon}_{\bar\nu}$ as 
the
graviton polarization tensor and the antisymmetric product as that of the
antisymmetric tensor field. It is obvious, the vertex operators for excited
closed string states can be derived by suitably generalizing (\ref{kltclosed})
as was achieved for the open string case. Moreover, the constraints arising 
from conformal invariance such as masslessness condition for the first excited
states are fulfilled. The  transversality conditions on polarization tensor
$\epsilon_{\mu\bar\nu}$ are translated to constraints on $\epsilon_{\mu}$
and ${\bar\epsilon}_{\bar\nu}$. Furthermore, the polarization tensors 
associated with excited massive states of closed string \cite{lo,jm2,jm3}
(for both compactified and noncompact strings) are constrained by requirements
of conformal invariance. Those conditions can be
 incorporated by imposing appropriate constraints on the set of polarization
vectors $\epsilon_{\mu_i}$ and ${\bar\epsilon}_{\bar\nu_i}$.   \\
Our goal is to suitably adopt the KLT approach to the vertex operators
associated with states arising due to compactification of a closed string. 
 In other words,
 excited  states living in uncompactified $\hat D$-dimensional theory, 
when compactified to lower dimension belongs to the representations of rotation
group $SO(D-1)$ (for massless case it is $SO(D-2)$) whereas these states
came from the representations of $SO({\hat D}-1)$ (correspondingly 
$({\hat D}-2)$ for massless case). Note that the moduli $G$ and $B$
transform as scalars under $D$-dimensional Lorentz transformations and spatial
rotations. Therefore, for an arbitrary excited states
which had Lorentz indices in $\hat D$-dimensions, will decompose into tensors,
vectors and scalars in $D$-dimensions \cite{jm2}.
 Consider compactification of $\hat D$
dimensional graviton; it decomposes into $D$-dimensional graviton, gauge bosons
and moduli. We focus the attention on the massless sector consisting of the 
moduli $G_{\alpha\bar\beta}, B_{\alpha\bar\beta}, 
A^{(1)\alpha}_{\mu} ~{\rm and}~
A^{(2)}_{\mu\alpha}$. Let us consider the KLM type vertex for the moduli
\be
\label{kltgb}
V^{\cal M}_{KLT}=:{\rm exp}[ik.X+i\epsilon_{\alpha}\partial Y^{\alpha}
+i{\bar\epsilon}_{\bar\beta}{\bar\partial}Y^{\bar\beta}]:
\ee
Remarks: (i) The scalars propagate in the spacetime manifold and therefore,
the plane wave part is of the form ${\rm exp}[ik.X]$. Thus throughout this
paper, we repeat, due to the Y-independence of plane waves 
in (\ref{kltgb})  we do not take into account
attributes of toroidal compactification such as presence of winding modes
etc. (ii) All the states are on-shell. As stated earlier, the vertex operators
of $G$ and $B$ are identified by taking symmetric and antisymmetric
combination of the product $\epsilon_{\alpha}{\bar\epsilon}_{\bar\beta}$. 
(iii) Note that while evaluating correlation function, starting from
$V^{\cal M}_{KLT}$ we shall have products of equal number of 
$\epsilon_{\alpha_i}$'s and ${\bar\epsilon}_{\bar\beta_i}$'s due to the 
level matching condition. The products of $\epsilon_{\alpha_i}$ will mutually
contract themselves due the $\delta^{\alpha_i\alpha_j}$ arising from 
contractions of $\partial Y^{\alpha_i}(z_i)$ and $\partial Y^{\alpha_j}(z_j)$,
$i\ne j$,  similar contractions appear from the antiholomorphic part as well.
The contraction, $<\partial Y^{\alpha_i}(z_i){\bar\partial} 
Y^{{\bar\beta}_n}({\bar z}_n)>=0$, consequently there are no contraction of the 
indices of type $\epsilon_{\alpha_i}$ and ${\bar\epsilon}_{\bar\beta_j}$ in 
evaluation of amplitudes. Lets us recall the $Z_2$ duality,
in the phase space approach: $P\leftrightarrow Y'$ under interchange
$\tau \leftrightarrow
\sigma$  of worldsheet coordinates. We note that, with
$P_{\pm}=(P\pm Y')$, $P_{\pm}\rightarrow \pm P_{\pm}$.
 Moreover, $O(d,d)$ vector ${\bf Z}$ can be decomposed as 
$\bf Z=({\bf Z}_+ +{\bf Z}_-)$. Where
\be
\label{z}
{\bf Z}_+={1\over 2}\pmatrix{P_{+} \cr P_{+} \cr},~~{\rm and}~~
{\bf Z_-}= {1\over 2} \pmatrix{P_{-} \cr  -P_{-} \cr}
\ee
Therefore, we  observe that the structure of the vertex operators for leading
Regge trajectories is intimately related 
with $O(d,d)$ symmetry from the phase space
perspective.\\
\subsection{Scattering of Moduli and Gauge Bosons for Compactified Stringy
States}
We have proposed the modified version of KLT vertex operators for the
compactified closed string. We shall employ this vertex operator to evaluate
N-point amplitudes for moduli $G$ and $B$ as well as for the gauge bosons
arising as a consequence of toroidal compactification.\\ 
The N-point amplitude extracted from $V^{\cal M}_{KLT}$ will have products of
$\epsilon$'s from the holomorphic sector and products of $\bar\epsilon$'s
from antiholomorphic sector. A careful examination of $V^{\cal}_{KLT}$
itself reveals an interesting fact. If we define $\epsilon_{\alpha_i}$
to transform as vector under, say, $O(d)_R$ and  $\partial Y^{\alpha_i}$ to be
also the  vectors of $O(d)_R$ then the second term in the exponential of
(\ref{kltgb}) is $O(d)_R$ invariant. Moreover, if I invoke similar
definition for the antiholomorphic sector i.e. declare 
$\{{\bar\epsilon}_{\bar\beta_i}, {\bar\partial}Y^{\bar\beta_i}\}$ 
to transform as
two sets of vectors under $O(d)_L$, then the third term in the exponential
is $O(d)_L$ invariant. Therefore, the polarizations $\epsilon_{\alpha_i}$
and ${\bar\epsilon}_{\bar\beta_i}$ transform separately under $O(d)_R$ and
$O(d)_R$ as ordained above. \\
Let us consider the N-point amplitude for
the moduli $G_{\alpha\bar\beta}$ and $B_{\alpha\bar\beta}$. We arrive
at the following general expression and then we shall discuss how to extract
the N-point amplitude alluding to remarks made earlier.
\bea
\label{ngbamp}
A^{(N)}_{G,B}=\int{\bf\Pi}_{i=1}^N d^2z_i~{\cal D}~{\bf\Pi}_{i>j}|z_i-z_j|^
{2k_i.k_j}{\rm exp}{\bf\Sigma}^N_{i>j}{{{\epsilon_i}.{\epsilon_j}}
\over{(z_i-z_j)^2}}
{\rm exp}{\bf\Sigma}^N_{i>j}{{{{\bar\epsilon}_i.{{\bar\epsilon}_j}}
\over{({\bar z}_i-{\bar z}_j)^2}}}
\eea
where ${\cal D}$ is the Koba-Nielsen factor: 
\be
\label{knfactor}
{\cal D}={{|z_a-z_b|^2|z_b-z_c|^2|z_c-z_a|^2}\over{d^2z_ad^2z_bd^2z_c} }
\ee
which gauge fixes the underlying $SL(2,C)$ invariance. The variables 
$\{z_a,z_b,z_c \}$ and $\{{\bar z}_a,{\bar z}_b,{\bar z}_c \}$  
can be chosen
arbitrarily. Therefore, there are only $(N-3)$ integrations over $\Pi d^2z_i$.
 Note that  
$\epsilon_i.\epsilon_j=\epsilon_{\alpha_i}\delta^{\alpha_i\alpha_j}\epsilon_{\alpha_j}$ and similar
definition applies for ${\bar\epsilon}_i.{\bar\epsilon}_j$. 
Since we have two independent products $ \Pi\epsilon_{\alpha_i}$
and $\Pi{\bar\epsilon}_{{\bar\beta}_i}$ and they contract only among 
themselves we can make the $O(d)_R$ and
$O(d)_L$ transformations on each of the products. 
Since these 'polarization vectors' contract amongst themselves,
as noted earlier, evidently, the N-point amplitude is T-duality invariant in 
the sense described above.\\
The vertex operators of the gauge bosons, which arise
 from dimensional reduction of the metric and antisymmetric
tensor fields,  are given by
\be
\label{kltamu}
V^{A}=:{\rm exp}[ik.X+i\epsilon_{\mu}\partial X^{\mu}+
i{\bar\epsilon}_{\bar\beta}^{(i)}{\bar\partial}Y^{\bar\beta}]:,~~i=1,2
\ee
in KLT formalism and we keep in mind  that in the expansion of the exponentials 
we retain the bilinear $\epsilon_{\mu}{\bar\epsilon}_{\bar\beta}$. 
The index of ${\bar\epsilon}^{(1)}_{\bar\beta}$, 
${\bar\beta}=1,2...d$ and this polarization is identified with 
$A^{(1){\bar\beta}}_{\mu}$. The other one  ${\bar\epsilon}^{(2)}_{\bar\beta}$, 
${\bar\beta}=1,2...d$ is polarization of gauge field $A^{(2)}_{\mu\bar\beta}$.
  In other words the third term
in the exponential of (\ref{kltamu}) ${\bar\epsilon}_i{\bar\partial}Y^i={\bar\epsilon}^{(1)}_{\bar\alpha}
{\bar\partial}Y^{\bar\alpha}+{\bar\epsilon}^{(2)}_{\bar\alpha}{\bar\partial}Y^{\bar\alpha}$.
 Now the index $\bar\alpha=1,2...d$. Moreover, the plane wave part
$e^{i.k.X}$ is invariant under $T$-duality and so is $\epsilon_{\mu}.\partial X^{\mu}$.  However, the photon polarization
$\epsilon_{\mu i}=\epsilon_{\mu}{\bar\epsilon}_i, ~\mu=0,1,...D-1,~ {\rm and}~ i=1,2..2d $ is factorized into products
of spacetime polarization vector and the internal polarization vector. Moreover, the T-duality group linearly acts on internal
polarizations, ${\bar\epsilon}_{\bar\alpha_i}$.
\\
The N-point amplitude for vector bosons assumes the following form  
\bea
\label{gaugeamp}
T^{(N)}_{A}\simeq &&\int d^2z_1d^2z_2...d^2z_N~{\cal D}~{\bf\Pi}_{i>j}
|z_i-z_j|^{2k_i.k_j} \nonumber\\&&
{\rm exp}[{\bf\Sigma} _{i>j}{{\epsilon_i.\epsilon_j}\over{(z_i-z_j)^2}}-
{\bf\Sigma}_{i\ne j}
{{\epsilon_i.k_j}\over{(z_i-z_j)}}]{\rm exp}[{\bf\Sigma}_{i\ne j}
{{{\bar\epsilon}_i.
{\bar\epsilon}_j}\over{({\bar z}_i-{\bar z}_j)^2}}]
\eea
Here $\cal D$ is the choice of the gauge fixing Koba-Nielsen measure.
The first term in the first exponential comes from contractions of
$\partial X^{\mu_i}(z_i)\partial X^{\mu_j}(z_j)$ 
and the second term in the first
exponential is due to contraction of plane wave and $\partial X^{\mu_i}(z_i)$.
Notice that there is no such term in the second exponential since the plane wave
part has no $Y^{\alpha}$ dependence. We need to pick up products of terms
like $\epsilon_i.\epsilon_j$ and $\epsilon_i.k_j$ from holomorphic side and
various pairwise contractions of ${\bar\epsilon}_i.{\bar\epsilon}_j$ from the
contractions of the antiholoporphic side. The objects like 
$\epsilon_i.\epsilon_j$ and $\epsilon_i.k_j$ are inert under T-duality 
transformations. Therefore, T-duality will only rotate the polarization
vectors $\{ {\bar\epsilon}_{\alpha_i} \}$.\\
In what follows, I shall present two illustrative examples to demonstrate
the operations of $T$-duality. \\
Let us consider a simple case where three
spatial dimensions are compactified, i.e. 
$d=3$. First we consider the four point
scattering scattering amplitude of the moduli $G_{\alpha\bar\beta}$.
The four vertex operators are given by
\be
\label{modulivertx}
V_i=\epsilon_{\alpha_i}{\bar\epsilon}_{{\bar\beta}_i}:{\rm exp}[ik.X(z,\bar z)]
\partial Y^{\alpha_i}(z_i){\bar\partial} Y^{{\bar\beta}_i}({\bar z}_i):,~~~
i=1,2,3,4
\ee
Thus the four point amplitude assumes the form
\be
\label{4pointg}
A^{(4)}\simeq\int {\bf\Pi}_{i=1}^4 d^2z_i~{\cal D}~
{\bf \Pi}_{ i>j}|z_i-z_j|^{2k_i.k_j}
{\rm exp}[{\bf\Sigma}_{i>j}{{\epsilon_i.\epsilon_j}\over{(z_i-z_j)^2}}]
{\rm exp}[{\bf\Sigma}_{i>j}{{{\bar\epsilon}_i.{\bar\epsilon}_j}
\over{({\bar z}_i-{\bar z}_j)^2}}]
\ee
We fix the three Koba-Nielesen variables as $ z_1=0,~z_3=1,~ z_4\rightarrow
\infty$ and correspondingly set ${\bar z}_1,{\bar z}_2,{\bar z}_3$ to those 
values. Thus we are left with an integral over $z_2$ and ${\bar z}_2$. 
We expand the two exponentials in power series and pick up on the term
which is a product 
$\epsilon_{\alpha_1}\epsilon_{\alpha_2}\epsilon_{\alpha_3}\epsilon_{\alpha_4}$
and ${\bar\epsilon}_{{\bar\beta}_1}{\bar\epsilon}_{{\bar\beta}_2}
{\bar\epsilon}_{{\bar\beta}_3}{\bar\epsilon}_{{\bar\beta}_4}$. We identify
$1-2$ as the incoming particles and $3-4$ as outgoing ones. The three
Mandelstam variables are
\be
\label{mandel}
s=-(k_1+k_2)^2,t=-(k_1+k_3)^2,u=-(k_1+k_4)^2,~
 {\rm and}~s+t+u=0
\ee
The energy momentum conservation lead to the condition: 
$(k_1+k_2+k_3+k_4)=0$.
It might be interesting to draw analogy with some familiar, well known
 problems
in quantum mechanics. We might imagine the product $\epsilon_{\alpha_1}
\epsilon_{\alpha_2}{\bar\epsilon}_{{\bar\beta}_1}{\bar\epsilon}_{{\bar\beta}_2}$
as the initial wave function and similarly 
$\epsilon_{\alpha_3}
\epsilon_{\alpha_4}{\bar\epsilon}_{{\bar\beta}_3}{\bar\epsilon}_{{\bar\beta}_4}$
as the final state wave function so far as the products of these 
polarizations are concerned. These tensors are in the internal space,
 unlike the polarization vectors of photons which transform as vectors under
Lorentz transformations. Moreover $\epsilon_{\alpha_i}$ is an $O(3)_R$ 
vector and ${\bar\epsilon}_{{\bar\beta}_i}$ is a $O(3)_L$ vector. Therefore,
the initial product tensor $\epsilon_{\alpha_1}\epsilon_{\alpha_2}$ can be
decomposed into a sum of symmetric traceless second rank tensor and a scalar.
The former is like a quadrupole operator in quantum mechanics when we carry
out similar decomposition for operator $x^ix^j$ (in radiative transitions in 
atomic and nuclear physics). Thus we express
\be
\label{qpole}
\epsilon_{\alpha_1}\epsilon_{\alpha_2}=(\epsilon_{\alpha_1}\epsilon_{\alpha_2}
-{1\over 3}\delta_{\alpha_1\alpha_2}\epsilon_1.\epsilon_2)+{1\over 3} 
\delta_{\alpha_1\alpha_2}\epsilon_1.\epsilon_2
\ee
We can carry out similar decomposition for the product
${\bar\epsilon}_{{\bar\beta}_1}{\bar\epsilon}_{{\bar\beta}_2}$.
We would like to define  'wave functions' for the  initial state and for
the final state. However, in order to adopt  compact notations and to 
bring out the
tensor structures under $O(3)_R$ and $O(3)_L$ rotations let us define
\be
\label{initialwf1}
Q^I_{\alpha_1\alpha_2}=(\epsilon_{\alpha_1}\epsilon_{\alpha_2}-{1\over 3}
\delta_{\alpha_1\alpha_2}\epsilon_1.\epsilon_2) ,~~~{\rm and}
~~~S^I_{\alpha_1\alpha_2}={1\over 3}
\delta_{\alpha_1\alpha_2}\epsilon_1.\epsilon_2
\ee
Correspondingly, the tensors of right mover's polarization are expressed as
\be
\label{initialwf2}
{\bar Q}^I_{{\bar\beta}_1{\bar\beta}_2} =       
({\bar\epsilon}_{{\bar\beta}_1}{\bar\epsilon}_{{\bar\beta}_2}-{1\over 3}
\delta_{{\bar\beta}_1{\bar\beta}_2}{\bar\epsilon}_1.{\bar\epsilon}_2)   
,~~~{\rm and}~~~
{\bar S}^I_{{\bar\beta}_1{\bar\beta}_2}={1\over 3}
\delta_{{\bar\beta}_1{\bar\beta}_2}{\bar\epsilon}_1.{\bar\epsilon}_2
\ee
For the final state wave functions we define
\be
\label{finalwf1}
Q^F_{\alpha_3\alpha_4}=(\epsilon_{\alpha_3}\epsilon_{\alpha_4}-{1\over 3}
\delta_{\alpha_3\alpha_4}\epsilon_3.\epsilon_4) ,~~~{\rm and}
~~S^F_{\alpha_3\alpha_4}={1\over 3}
\delta_{\alpha_3\alpha_4}\epsilon_3.\epsilon_4
\ee
and
\be
\label{finalwf2}
{\bar Q}^F_{{\bar\beta}_3{\bar\beta}_4} =
({\bar\epsilon}_{{\bar\beta}_3}{\bar\epsilon}_{{\bar\beta}_4}-{1\over 3}
\delta_{{\bar\beta}_3{\bar\beta}_4}{\bar\epsilon}_3.{\bar\epsilon}_4)
,~~~{\rm and}~~~
{\bar S}^F_{{\bar\beta}_3{\bar\beta}_4}={1\over 3}
\delta_{{\bar\beta}_3{\bar\beta}_4}{\bar\epsilon}_3.{\bar\epsilon}_4
\ee
The the four point scattering amplitude for the moduli takes the following
form after we have fixed the three Koba-Nielsen variables and therefore,
left with one integration
\bea
\label{compactamp}
A^{(4)}(s,u)\simeq=\int d^2z_2|z|^{2k_1.k_2}|1-z_2|^{2k_2.k_3}
({\bf\Sigma}_{a=1}^3 T^a_R) ({\bf\Sigma}_{a=1}^3 {\bar T}^b_L)
\eea
thus the product results in a  total of nine terms 
(there are nine terms in the
integral (\ref{compactamp})). The terms $T^i_R$ and ${\bar T}^i_L$ are
\be
\label{tr1}
T^1_R={1\over{(z_2)^2}}\bigg(Q^I_{\alpha_1\alpha_2}+S^I_{\alpha_1\alpha_2}\bigg)
\bigg(Q^F_{\alpha_3\alpha_4}+S^F_{\alpha_3\alpha_4}\bigg)
\delta^{{\alpha_1}{\alpha_2}}\delta^{{\alpha_3}{\alpha_4}}
\ee
\be
\label{tr2}
T^2_R={1\over{(z_2)^2}}\bigg(Q^I_{\alpha_1\alpha_2}+S^I_{\alpha_1\alpha_2}\bigg)
\bigg(Q^F_{\alpha_3\alpha_4}+S^F_{\alpha_3\alpha_4}\bigg)
\delta^{{\alpha_1}{\alpha_3}}\delta^{{\alpha_2}{\alpha_4}}
\ee
\be
\label{tr3}
T^3_R={1\over{(1-z_2)^2}}\bigg(Q^I_{\alpha_1\alpha_2}+
S^I_{\alpha_1\alpha_2}\bigg)
\bigg(Q^F_{\alpha_3\alpha_4}+S^F_{\alpha_3\alpha_4}\bigg)
\delta^{{\alpha_1}{\alpha_4}}\delta^{{\alpha_2}{\alpha_3}}
\ee
Notice the following: (i) $T^1_R,~T^2_R~{\rm and}~ T^3_R$ have factors of
$(z_2)^{-2},1, (1-z_2)^{-2}$ respectively multiplying them.(ii) Although we
have the same products like $[(Q^I+S^I)_{{\alpha_1}{\alpha_2}}][
(Q^F+S^F)_{{\alpha_3}{\alpha_4}}]$ in the expressions for  
$T^1_R,~T^2_R~{\rm and}~ T^3_R$ the  Kronekar $\delta$'s contracting them is
different. Thus we have trace of the products of these tensors. Of course,
it is natural since these products are expected to be eventually $O(3)_R$
invariants. Since $Q^I$ and $Q^F$ are traceless, $T^1_R=
({\rm Tr}S^I)({\rm Tr}S^I)$. The other two terms $T^2_R$ and $T^3_R$ have the
same structure, although the former has $1$ as a coefficient and latter has
$(1-z_2)^{-2}$. This term, in the expressions for $T^2_R$ and $T^3_R$, is
\be
\label{t2t3}
{\rm Tr}\bigg(Q^IQ^F+Q^IS^F+S^IQ^F+S^IS^F \bigg)
\ee
The expressions for ${\bar T}^i_L$ are
\be
\label{tbar1}
{\bar T}^1_L={1\over{{{\bar z}_2}^2}}({\rm Tr}{\bar S^I}) ({\rm Tr}{\bar S^F})
\ee
\be
\label{tbar2}
{\bar T}^2_L={\rm Tr}\bigg({\bar Q}^I{\bar Q}^F+{\bar Q}^I{\bar S}^F+
{\bar S}^I{\bar Q}^F+{\bar S}^I{\bar S}^F \bigg)   
\ee
and
\be
\label{tbar3}  
{\bar T}^3_L={1\over{(1-{\bar z}_2)^2}}   
{\rm Tr}\bigg({\bar Q}^I{\bar Q}^F+{\bar Q}^I{\bar S}^F+
{\bar S}^I{\bar Q}^F+{\bar S}^I{\bar S}^F \bigg)
\ee
As expected $\{T^i_L \}$ are $O(d)_L$ invariant. The products
$({\bf\Sigma}_{a=1}^3 T^a_R)({\bf\Sigma}_{a=1}^3 {\bar T}^a_L)$ have 
altogether nine terms. These 
integrals can be evaluated using the standard methods. We draw attention to the
fact that $2k_1.k_2=-s$ and $2k_1.k_3=-u$ since we are considering scattering
of massless particles. The integral
\be
\label{vs}
\int d^2|z_2|^{-s}|1-z_2|^{-u}= 2\pi{{\Gamma(1-{s\over 2})\Gamma(1-{u\over 2})
\Gamma(-1-{s\over 2}-{u\over 2}) }
\over{\Gamma({s\over 2})\Gamma({u\over 2})\Gamma(2-{s\over 2}-{u\over 2}) }    }
\ee
There are other terms like $z_2^{-2}, (1-z_2)^{-2}...$ and so on.
which multiply this integrand.  Each of these integrations 
 can be handled by using table
integrals for gamma functions (see \cite{klt} for discussion on evaluating
these type of integrals, especially the appendix). We mention in 
passing that the T-duality invariance of the four point amplitude is 
manifest in the procedure we proposed above.\\
Let us consider the four point function involving gauge bosons. We could
compute four gauge boson amplitude; however, in order to bring out the
general features, it will suffice to consider scattering of a gauge boson
$A^{{\bar\alpha_i}}_{\mu}$ from a tachyon where $i=1,2,3,4,5,6$. The
initial and final tachyon momenta are $k_1$ and $k_3$ and those of the
two gauge bosons are $k_3$ and $k_4$. The amplitude is
\be
\label{gtamp}
T^{(4)}=\int{\bf\Pi}_{i=1}^4 d^2z_i{\cal D}{\bf\Pi}_{i>j}|z_i-z_j|^{2k_i.k_j}
{\rm exp}[{{\epsilon_2.\epsilon_4}\over{(z_2-z_4)^2}}-{\bf\Sigma}_{i\ne j}
{{\epsilon_i.k_j}\over{(z_i-z_j)}}]{\rm exp}
[{{{\bar\epsilon}_2.{\bar\epsilon}_4}\over{({\bar z}_2-{\bar z}_4)^2}}]
\ee
after we expand the exponentials and collect the relevant terms.  These terms
are of the form $\epsilon_2.\epsilon_4$ times
 ${\bar\epsilon}_2.{\bar\epsilon_4}$ that will appear in four point amplitude
since we have only two photons.
${\cal D}$ is the Koba-Nielsen factor and we choose it to be the same as
before and therefore, the above expression (\ref{gtamp}) consists of an
integral over $z_2$ and ${\bar z}_2$. 
The terms like $\epsilon_1.\epsilon_2$ and $\epsilon_i.k_j$ have
passive transformations under  T-duality.
 Therefore, we need pay attention to this piece. The term of interests to us
is ${\bar\epsilon}_2.{\bar\epsilon}_4$ and it is automatically $T$-duality
invariant. However, we shall discuss how we can generate new scattering
amplitudes from (\ref{gtamp}). \\
It is interesting to note that the four point function for gauge 
bosons (Abelian)
can be evaluated efficiently in this formalism. We start from the general
expression and the outline the prescription  (amplitude is $A^{(4)}_{gauge}$)
\be
\label{fourgauge}
A^{(4)}_{gauge}\simeq\int{\bf\Pi}_{i=1}^4d^2z_i~{\cal D}~{\bf\Pi}_{i>j}
|z_i-z_j|^{2k_i.k_j}{\cal E}_1{\cal E}_2
\ee
where ${\cal E}_1$ and ${\cal E}_2$ stand for two exponentials defined below
\be
\label{exp1}
{\cal E}_1=
{\rm exp}[{\bf\Sigma}_{i>j}{{\epsilon_i.\epsilon_j}\over{(z_i-z_j)^2}}-
{\bf\Sigma}_{i\ne j}{{\epsilon_i.k_j}\over{(z_i-z_j)}}]
\ee
and
\be
\label{exp2}
{\cal E}_2=
{\rm exp}[
{\bf\Sigma}_{i>j}
{{{\bar\epsilon}_i.{\bar\epsilon}_j}\over{({\bar z}_i-{\bar z}_j)^2}} ]
\ee
The four point function will have one integration left  since
we choose $\{z_1,z_3,z_4 \}$ to be the Koba-Nielsen variable and assign them
the values as before (similarly we assign values to the other complex
conjugates as before). We discuss the structure of the amplitude.
We have  polarizations $\{\epsilon_{\mu_i}\}$ which are spacetime vectors
and we have $\{{\bar\epsilon}_i \}$ which are 'polarizations' coming from
internal directions. The latter will mutually contract among themselves.
The set ${\epsilon_{\mu_i}}$ will not only contract among themselves but there
will be products like $(\epsilon_i.\epsilon_j)(\epsilon_m.k_n)$ and so on.
The essential points to note are there have to be products of four
$\epsilon_{\mu_i}$ which are contracted amongst themselves or with $k^{\mu_j}$
keeping in mind that $\epsilon_{\mu_i}.k^{\mu_i}=0$ (no sum over $\mu_i$).
Such an amplitude has been computed before in several ways. We intend to 
present arguments which will bring out the essential features of 
$A^{(4)}_{gauge}$ in this regard. Let us look at ${\cal E}_2$ first.
The we have to retain up to the quadratic term in the expansion of the
 exponential. The terms of our interests are of the form 
${\bar\epsilon}_{{\bar\alpha}_i}.{\bar\epsilon}_{{\bar\alpha}_j}
{\bar\epsilon}_{{\bar\alpha}_m}.{\bar\epsilon}_{{\bar\alpha}_n}$ where
$i,j,m,n,~{\rm take~ values}~1,2,3,4$ they have to be different in the
products. If we go back to the 4-point amplitude for scattering of moduli
and in the present case 
identify ${\bar\epsilon}_{{\bar\alpha}_1} {\bar\alpha}_{{\bar\alpha}_2}$ as
the initial 'wave function' and 
${\bar\epsilon}_{{\bar\alpha}_3} {\bar\alpha}_{{\bar\alpha}_4}$ as the final
wave function then the arguments of (\ref{t2t3}) to (\ref{tbar3}) go through. Thus
the T-duality invariance of $A^{(4)}_{gauge}$ is ensured.
Let us discuss the terms coming from expansion of ${\cal E}_1$.
Here we have to first retain up to quadratic term in the expansion and some
interference terms coming from the cubic part and terms coming from
quartic  part. (I) We note the presence of the products
 of the type 
$(\epsilon_i.\epsilon_j)(\epsilon_m.\epsilon_n),~i\ne j, m\ne n$ and $i,j,m,n$
are all different. These originate from the quadratic terms in the
expansion of the exponential. (II) The other type of terms are 
$(\epsilon_i.\epsilon_j) 
(\epsilon_m.k_l)(\epsilon_n.k_s)$. Such types of terms originate from
the cubic term in expansion of exponential - interference of two 
${\bar \epsilon}.k$ with $\epsilon.\epsilon$ (all $\epsilon$ indices are
to be different).  Note that we should have 
all four polarization vector's indices are different.
 (III) There exist
another type of term which are like $\epsilon_i .k^j$ where four of them
will occur. Such types of terms will come from the fourth power in the
expansion of the first exponential. In order to make this argument transparent,
let us look at one combination, 
(there are more such terms)
\be
\label{exp3}
\bigg({{\epsilon_1.k_2}\over{(z_1-z_2)}}+{{\epsilon_2.k_1}\over{(z_2-z_1)}}
+{{\epsilon_3.k_4}\over{(z_3-z_4)}}+{{\epsilon_4.k_3}\over{(z_4-z_3)}}\bigg)^2
\ee
There is an interference term (suppressing the difference $(z_i-z_j)$ which 
will all appear as products in the denominator):
\be
\label{inter1}
{\epsilon_1.k_2}{\epsilon_2.k_1}{\epsilon_3.k_4}{\epsilon_4.k_3}
\ee
apart from a numerical factor. There will be several combinations
of $\epsilon.k$  in such products.\\
There are a few  points to be made:\\
 (i) So far as $T$-duality
transformation is concerned, these are manifestly invariant.\\
(ii) We should bring in the $\alpha'$ dependence to see the other aspect.
Notice that in ${\cal E}_1$ each term in multiplied by $\alpha'$ since the 
first one comes from correlation of pair like
$\partial X^{\mu_i}\partial X^{\mu_j}=-{{\alpha'}\over 2}
{{\delta^{\alpha_1\alpha_2}}\over{(z_i-z_j)^2}}$ 

and the other
one
comes from contraction of $\partial X^{\mu_i}:e^{k_j.X_j}$. This also has
a factor of $\alpha'$. Therefore, when we expand the exponential, we have
 different powers of $\alpha'$. Of course this is expected on dimensional 
ground. 
Note that this is already seen in the three point vertex of graviton
\cite{book2}. There are
terms which are linear in momenta with suitable metric multiplied to it
( in the sense when we look at contributions of the holomorphic parts). \\
All these amplitudes are calculated through applications of known conformal
field theory techniques.
 However, the KLT prescription is more economical at the tree
level.\\ 
Now we present some applications of $T$-duality transformations to 
demonstrate how  another four point function will be generated from a given
one. Let us consider the case of scattering of moduli (\ref{compactamp}).  
We choose 
\be
\label{ex1} 
\epsilon_{\alpha}=\pmatrix{1 \cr 0\cr 1\cr},~{\rm and}~~
{\bar\epsilon}_{\bar\beta}=\pmatrix{1 \cr 0\cr -1\cr}
\ee
The corresponding moduli $G$ and $B$ are
\be
\label{oldgb}
G_{\alpha{\bar\beta}}=\pmatrix{1 & 0 & 0 \cr 0 & 0 & 0 \cr 0 & 0 & -1\cr},
~~{\rm and}~~
B_{\alpha{\bar\beta}}=\pmatrix{0 & 0 & -1\cr 0 & 0 &  0\cr -1 & 0 &0 \cr}
\ee
Now consider a simple $T$-duality transformation where we rotate by $O(3)_R$
on the $2-3$ plane and denote this operation by $R$. The other rotation is
$S$,
 the 
$O(3)_L$ and  $S$ is also a rotation on $2-3$ plane. 
We identify $S=R^T$. We choose
\be
\label{r-rotation}
R=\pmatrix{1 & 0 & 0\cr 0 & cos\theta & sin\theta \cr 0 &-sin\theta & cos\theta
\cr},~~~ {\rm  and }~~~
S=\pmatrix{1 & 0 & 0\cr 0 & cos\theta & -sin\theta \cr 0 & sin\theta & cos\theta
\cr}
\ee
Let the rotation angle be  $\theta={\pi\over 4}$ for simplicity, 
so that $cos\theta=sin\theta
={1\over{\sqrt 2}}$. According to our prescriptions the transformed vectors are
\be
\label{rvectors}
\epsilon'_{\alpha}=\pmatrix{1 \cr{1\over{\sqrt 2}}\cr{1\over{\sqrt 2}} \cr},~~
{\rm and} ~~~{\bar\epsilon}'_{\bar\beta}=\pmatrix{1 \cr{1\over{\sqrt 2}}\cr 
-{1\over{\sqrt 2}} \cr}
\ee
and the two transformed backgrounds are
\be
\label{primedgb}
G'=\pmatrix{1 & {1\over{\sqrt 2}} & 0 \cr 
{1\over{\sqrt 2}}  &  {1\over 2} & 0\cr
0 & 0 & -{1\over 2}\cr},~~~ {\rm and}~~~
B'=\pmatrix{0 & 0 & -{1\over{\sqrt 2}}\cr 0 & 0 & -{1\over 2} \cr
       {1\over{\sqrt 2}}   &  {1\over 2} & 0\cr  }
\ee
Thus this simple example brings out the essential features of $T$-duality
transformation in this approach. For a most general rotation, the $O(3)$
group is parametrized by three Euler rotation matrices. We are free to choose
them for $O(3)_R$ and $O(3)_L$. Therefore, starting from very simple 
vectors $\epsilon_{\alpha}$ and ${\bar\epsilon}_{\bar\beta}$ as in 
(\ref{ex1}), we can generate very complicated form of these vectors.
Moreover, if we have the four point amplitude for these configurations,
we can connect the amplitude to the one which has 
more complicated configurations of polarization vectors.\\
Let us turn our attention to the gauge  boson and tachyon 
scattering  amplitude (\ref{gtamp}).
Here we shall focus on one term ${\bar\epsilon_2}.{\bar\epsilon_4}$. Recall
that now $\bar\epsilon$ is a six component vector; three of them coming 
from  ${\bar\epsilon}^{(1)}_{\bar\beta},~{\bar\beta}=1,2,3$ and the other
three coming from ${\bar\epsilon}^{(2)}_{\bar\beta},~{\bar\beta}=1,2,3$. Let us remind ourselves
that $A^{(1){\bar\beta}}_{\mu}$ and $A^{(2)}_{\mu{\bar\beta}}$ transform as
a doublet under $O(3,3)$: ${\cal A}_{\mu}\rightarrow \Omega {\cal A}_{\mu}$
According to Sen's prescription
\be
\label{s+r}
\pmatrix{A^{(1)}_{\mu} \cr A^{(2)}_{\mu}\cr} \rightarrow
{1\over 2}\pmatrix{R+S & R-S \cr R-S & R+S \cr}
\pmatrix {A^{(1)}_{\mu} \cr A^{(2)}_{\mu}\cr}
\ee
Therefore, the two sets of polarizations i.e. ${\bar\epsilon}^{(1)}$ and
${\bar\epsilon}^{(2)}$, we are dealing with, will transform according to
(\ref{s+r}). Thus if we had a configuration where the photon originated
from compactification of $\hat D$-dimensional graviton, then through the 
$O(3)\otimes O(3)$ transformation we  will relate this amplitude to
scattering of gauge bosons which are admixtures of $A^{(1)}_{\mu}$ and
$A^{(2)}_{\mu}$. This example illustrates the central point. If we have
amplitude with four gauge bosons, we know how to extract the $T$-duality
transformation part of it using the above steps.
As an example, let us start with a configuration for gauge boson
and tachyon scattering where we have only the
gauge field $A^{(1)\bar\beta}_{\mu}$ and we set $A^{(2)}_{\mu\bar\beta}=0$.
Then under $O(3)_R\otimes O(3)_L$ rotations: $A'^{(1)\bar\beta}_{\mu}
\rightarrow (R+S) A^{(1)\bar\beta}_{\mu}$ and 
$A'^{(2)}_{\mu\bar\beta}=(R-S)_{\bar\beta\bar\alpha}A^{(1)\bar\alpha}_{\mu}$. 
Thus we generate 
a scattering amplitude involving both the gauge fields.
We can interpret this result as follows. If we start from gravi-photon (
gauge boson originating from dimensional reduction of graviton in
$\hat D$-dimensions) and tachyon amplitude then through above duality
rotation it gets related to state which is admixture of gravi-photon and
axi-photon (gauge boson originating from reduction of $\hat B$-field.
\subsection{Scattering of nonabelian Bosons in KLT Formalism}
In this short subsection, we propose a generalization of the KLT approach to
construct vertex operators for nonabelian states in certain compactified 
schemes. The most familiar example is the heterotic string. The Yang-Mills
super multiplet appears in the massless sector together with 
the $N=1$ supergravity multiplet. There are excited massive states \cite{het1}
belonging to the representations of the chosen gauge group. Therefore,
it is of interests to construct vertex operators for such states which carry
the nonabelian charges. We  briefly recapitulate below how the vector boson
vertex operator was constructed in the bosonic coordinate representation when
compact coordinates are along the torii and the canonical momenta 
are quantized with further restrictions \cite{het2}.
 Here we present another approach 
which is useful to compute tree level amplitudes. We recall that there are gauge
bosons in the adjoint representation of 
 $SO(32)$ or $E_8\otimes E_8$ in heterotic
string theory in super Yang-Mills sector. These nonabelian vector states
appear if the lattice is self-dual and even. In order to fulfill
this requirements, the number of compactified dimensions have to be 
multiples of $8$. One of the most attractive features of heterotic string
theory is that in, $D=10$, these are the only possible gauge groups appear 
since $16$ of the twenty six coordinates of the bosonic sector are
toroidally compactified. It is remarkable that these two gauge groups 
are precisely the admissible gauge groups for cancellation of anomalies
discovered by Green and Schwarz \cite{gs}.
The scattering of these nonabelian gauge bosons were studied in \cite{het2}.
The vertex operators not only have the usual spacetime plane wave factor 
 but also
 a certain term ${\rm exp}[2iP_I.Y^I]$ where $P_I$ are the momenta are along
compact direction,  they are quantized and satisfy $(P^I)^2=2$. 
Note the appearance of $2P^I$
in vertex operator: this is generator of translation in the internal space.
The emission vertex for the charged gauge field needs another factor - the 
operator cocycle ${\hat C}(K^I)$ and its action on a state of momentum $P^I$
gives the two cocycle $\epsilon(P,K)$
\be
\label{cocy}
{\hat C}(K)|P^I>=\epsilon(P,K)|P^I>
\ee
Moreover, ${\hat C}(K){\hat C}(L)=\epsilon(K,L){\hat C}(K+L)$ and the two
cocycle condition is 
$\epsilon(K,L)\epsilon(K+L,M)=\epsilon(L,M)\epsilon(K,L+M)$. It is possible
to choose the cocycles such that (i) 
$\epsilon(K,L+M)=\epsilon(K,L)\epsilon(K,M)$.(ii) They satisfy
$\epsilon(K,L)\epsilon(L,K)=(-1)^{K.L}$, $\epsilon(K,0)=-\epsilon(K,-K)=1$, 
for $K^2=L^2=(K+L)^2=2$, it coincides with structure constants of the
group. The three point and four point Yang-Mills amplitude have been 
evaluated long since \cite{het2}. The structure constants of the Yang-Mills 
group is identified to be 
$f^{K_1K_2K_3}=\epsilon(K_2,K_3)\delta_{(K_1-K_2-K_3)}$. 
Indeed, Kawai,Llewellen and Tye \cite{klt} adopted this form of 
vertex operator and used the properties mentioned above to evaluate the
gauge boson scattering amplitudes in their reformulation. 
\\
Let us consider a related string compactification of a closed bosonic string.
The left moving and right moving sectors are independent and these coordinates
can be compactified separately. Let us toroidally 
compactify $16$ coordinates of left moving sector and compactify same number
of coordinates in right moving sector as well. The ground state of the theory
is tachyonic. The $16$ compact coordinates in left and right moving 
 sector can be fermionized
to give 32 Weyl Majorana fermions in each sector. If we chose NS-NS boundary
conditions for all left and right moving fermions then the gauge group is
$SO(32)\otimes SO(32)$. The massless spectrum is quite interesting. This
compactified string 
has massless graviton, antisymmetric tensor field and dilaton.
In addition there are two copies of gauge bosons in the adjoint of the 
two $SO(32)$ groups coming from left and right movers. Moreover, the theory
has massless scalars transforming as $(496,496)$. \\
Let us consider the vertex operators for the gauge bosons in the compactified 
closed bosonic string
\be
\label{aleft}
V^{(L)}= :A^L_{\mu ij}(X){\bar\partial}X^{\mu}(\bar z)\psi^i(z)\psi^j(z):
\ee
and the other vertex operator is
\be
\label{aright}
V^{(R)}=:A^R_{\mu ij}(X)\partial X^{\mu}(z){\tilde \psi}^i(\bar z)
{\tilde \psi}^j(\bar z):
 \ee
Notice that each of the gauge bosons are in the adjoint of their $SO(32)$.
In the plane wave approximation we express the two vertex operators as gauge 
\be
\label{planewv1}
A^L_{\mu ij}=\epsilon^a_{\mu}(T^a)_{ij}:{\rm exp}[ik.X]
{\bar\partial}X^{\mu}(\bar z)
\psi^i(z)\psi^j(z):,
\ee
  and
\be
\label{planewv2}
A^R_{\mu ij}={\bar\epsilon}^a_{\mu}({\tilde T}^a)_{ij}:{\rm exp}[ik.X]
{\partial}X^{\mu}( z)
{\tilde\psi}^i({\bar z}){\tilde\psi}^j({\bar z}):
\ee
The two generators $T^a$ and ${\tilde T}^a$ in the vector representation
of the groups. The correlation functions for gauge boson amplitudes are
evaluated through these vertex operators. Let us generalize the KLT vertex
operator to the case of nonabelian gauge boson emission.
\be
\label{vertexna}
V_{gauge}=:{\rm exp}[ik.X+i{\bar\epsilon}_{\mu}{\bar\partial}X^{\mu}({\bar z})
+i\epsilon_{\mu}\partial X^{\mu}+i\epsilon^aT^a_{ij}\psi^i(z)\psi^j(z)+
i{\bar\epsilon}^a{\tilde T}^a_{ij}{\tilde\psi}^i(\bar z){\tilde\psi}^j(\bar z)]:
\ee
Notice the following features: (i) When we expand the exponential in powers
of the 'polarizations' we shall recover the vertex operators in the massless
sector.\\
(ii) The bilinear term $\epsilon_{\mu}{\bar\epsilon}_{\nu}$ will give
the vertex operators from graviton and antisymmetric tensor as mentioned
earlier.\\
(iii) The two bilinears ${\bar\epsilon}_{\nu}{\bar\epsilon}^a{\tilde T}^a$ and
${\epsilon}_{\mu}\epsilon^aT^a$ will give the two vertex operators for gauge
bosons once we identify ${\bar\epsilon}^a_{\mu}({\tilde T}^a)_{ij}$ and 
$\epsilon^a_{\mu}(T^a)_{ij}$ with the first and second term respectively. \\
This generalization of KLT vertex will make calculation of scattering
 amplitudes quite efficient. As an example consider the three point function
for gauge bosons $A^R$. 
The general structure of the three point function is 
\be
\label{threeglue}
\epsilon_{\mu_1 a}\epsilon_{\mu_2 b}\epsilon_{\mu_3 c}
T^{{\mu_1\mu_2\mu_3}abc}
\ee
Note that $T^{{\mu_1\mu_2\mu_3}abc}$ factorizes into a product $f^{abc}$,
the structure constant and a tensor with spacetime indices (see equation
below).
Since the three Koba-Nielsen variables are fixed, there is no integration
to be done. This three point function was already derived in \cite{het2}.
If we adopt the vertex operator proposed above (\ref{vertexna}) then tree 
level calculation becomes simpler. The first point to observe that polarization
vector $\epsilon^a_{\mu} $ factorizes as $\epsilon^a_{\mu}=\epsilon_{\mu}
{\bar\epsilon^a}$  for this case (i.e. gauge boson $A^R$). The second point is
is that the we have to expand the exponentials and collect the products
of the type: $\epsilon_{\mu_1}\epsilon_{\mu_2}\epsilon_{\mu_3}
{\bar\epsilon}^a{\bar\epsilon}^b{\bar\epsilon}^c$. Next we compute the 
correlation functions of right movers and left movers. Thus 
the product terms are
\be
\label{threeg}
f^{abc}\bigg(g^{\mu_1\mu_2}k^{\mu_3}_{12}+g^{\mu_2\mu_3}k^{\mu_1}_{23}+
g^{\mu_3\mu_1}k^{\mu_2}_{31} \bigg)
\ee
where $g^{\mu_i\mu_j}$ is the flat spacetime metric and $k^{\mu}_{ij}=
(k_i-k_j)^{\mu}$.  The form of (\ref{threeg}) can be cast in a differently
using energy momentum conservation condition $(k_1+k_2+k_3)=0$ and that
$\epsilon_{\mu_i}.k^{\mu_i}=0$ (no sum over $i$).
The structure constant comes from contractions of the fermions 
appearing in expansion of exponential. The spacetime tensor structure,
as is well known comes from various contractions involving spacetime 
string coordinates.
We have used 
$<{\tilde \psi}^i({\bar z}_i){\tilde\psi}^j({\bar z}_j)>\simeq
{{\delta^{ij}}\over{({\bar z}_i-{\bar z}_j)}}$.  We believe this 
prescription can be used to evaluate higher point amplitudes.

\section{Summary and Conclusions}
We set out to investigate $T$-duality transformation properties of 
scattering amplitudes. It is accepted that $T$-duality is a symmetry of
compactified string theory and the S-matrix is expected to be invariant.
We need to construct vertex operators in order to evaluate the amplitudes.
This is achieved in the weak field approximation. For closed bosonic string,
compactified on a d-dimensional torii the spectrum 
contains the moduli, gauge bosons,
graviton, antisymmetric tensor and dilaton, in its massless
sector. The moduli can be cast in an
form that transforms as an adjoint under the $O(d,d)$ transformation. 
However, it is not very convenient to implement $O(d,d)$ transformations
when the weak field expansion is made around the trivial background. We
have followed Sen's \cite{sen1} argument and identified 
$O(d)\otimes O(d) \in O(d,d) $ as
the duality group for our purpose.
 These choice of this duality group is very appropriate to study
transformation properties of the vertex operators. Furthermore, we employed
a suitably modified version the KLT \cite{klt} formalism to construct 
vertex operators for the moduli. This formulation has the advantage that
the polarization tensors of the moduli factorize when we construct the
vertex operators. Thus the $O(d)\otimes O(d)$ transformation properties
of the amplitudes become simple and transparent. In particular, the 
N-point amplitude for moduli $G$ and $B$ can be expressed in a compact form.
The Abelian gauge bosons, resulting from the toroidal compactification,
 transform linearly under $O(d,d)$ group. In the first place, the polarization
vector of these gauge bosons are shown to factorize by adopting the KLT
formulation. The N-point amplitude for  moduli and gauge bosons 
is demonstrated to be duality invariant. 
 The present investigation substantiated the arguments
of \cite{hsz} through explicit calculations that the S-matrix is indeed
T-duality invariant.\\
We presented two illustrative examples. We considered a case where three
string coordinates are compactified on three torus, $T^3$. First we considered
the four point function of the moduli. The $O(3)_R\otimes O(3)_L$ 
invariance of this amplitude can be explicitly verified by adopting
arguments used in quantum mechanics. The initial state, as far as the products
of polarizations are concerned, are decomposed into sum of irreducible 
representations of $O(3)_L$ and similarly as sum of irreducible 
representations of $O(3)_R$. Thus the initial function (in the space of
polarizations) is a director product of tensors. The same procedure
can be followed for the final state wave functions. These tensors
are contracted in various way by the metrics resulting from contractions of
string coordinates in the vertex operators. Therefore, the four point
amplitude is expressed in a manifestly duality invariant form.
This argument can be extended to computation of four point functions where
 $d$-coordinates are compact. Thus the initial states consisting
of left movers will be decomposed to sum of irreducible representations
of $O(d)_L$ and similarly, the initial wave function consisting of right
movers will be expressed as sum of irreducible representations of
$O(d)_R$. The same procedure will be applicable to the final state
wave functions as well. Therefore, all the four point amplitudes, in the
 massless sector can be demonstrated to be duality invariant.\\
We propose vertex operators, based on KLT formulation, for scattering of
nonabelian stringy states. This prescription is economical for the 
computation of $S$-matrix elements.
We explicitly evaluate three gauge boson vertex which agrees with known
results. This technique might be useful for computation of excited stringy
states which carry 'color' gauge charges.
\\
In summary, we have demonstrated that the amplitudes constructed for the
massless states of compactified closed bosonic string can be expressed
in manifestly $T$-duality invariant form. This is facilitated efficiently
through the
introduction of the vertex operators of the massless states. These vertex
operators are modified versions of those introduced by KLT. Moreover,
we proposed vertex operators  to evaluate amplitudes for scattering
of nonabelian gauge bosons. We noted earlier that such nonabelian gauge boson
appear as massless states of $D=10$  heterotic string as well as
in compactification of closed bosonic string as long as the internal moments
fulfill the criterion mentioned already. Moreover, there will be nonabelian
excited states. The scattering of these states from gauge bosons can be
evaluated by using the vertex operator introduced here and the generalization
there of.

\bigskip

\noindent{\bf Acknowledgments:}
I am thankful to Hikaru Kawai for very valuable discussions and
for sharing his deep insights of KLT formalism and for 
very useful correspondence. I enjoyed
his very warm hospitality at Kyoto. I would like to thank Yoichi Kazama for
discussions on duality symmetry of scattering amplitudes. The gracious
and warm hospitality of Satoshi Ito at KEK is gratefully acknowledged.

\end{document}